\documentclass[]{raa}            
\usepackage{graphicx,times}
\usepackage{natbib}

\providecommand{\bm}[1]{\mbox{\boldmath$#1$\unboldmath}}

\begin{document}

   \title{Hydrodynamical wind on magnetized Accretion Flows with Convection
}

   \volnopage{Vol.0 (200x) No.0, 000--000}      
   \setcounter{page}{1}          

   \author{S. Abbassi
      \inst{1, 2}
   \and A. Mosallanezhad
      \inst{1,3}
   }

   \institute{School of Physics, Damghan University, P.O.Box 36715-364, Damghan, Iran\\
        \and
             School of Astronomy, Institute for Research in Fundamental Sciences, P.O.Box 19395-5531, Tehran, Iran; {\it abbassi@ipm.ir\\
        \and
             Center for Excellence in Astronomy \& Astrophysics (CEAA - RIAAM) - Maragha, IRAN, P. O. Box: 55134 - 441,\\}
   }
   \date{Received~~2009 month day; accepted~~2009~~month day}

\abstract{We present self-similar solutions for radiation inefficiently accretion flows (RIAF) around black holes in the presence of outflow and global magnetic field. The influence of outflow is taken into account by adopting a radius dependent of mass accretion rate $ \dot{M} = \dot{M}_{0}(r/r_{0})^{s} $ with $ s > 0 $. Also we consider convection through a mixing length formalism to calculate convection parameter $ \alpha_{con} $. Our numerical results show that by increasing all components of
magnetic field , the surface density and rotational velocity increase although the
sound speed and radial infall velocity of the disc decrease. Also we have found out that the existence of wind will lead to reduction of surface density as well as rotational velocity. Moreover the radial velocity, sound speed, advection parameter and the vertical thickness of the disc will increase when outflow becomes important in the radiation inefficiently accretion flow.
}

   \authorrunning{S. Abbassi, A. Mosallanezhad}            
   \titlerunning{Hydrodynamical wind on magnetized Accretion Flows with Convection}  

   \maketitle

%
%
\section{Introduction}           
\label{sect:intro}

Observations of accreting black holes of different mass show impressive similarities of data which point at
identical physical process in the accretion flow. Also, black hole accretion discs demonstrate a great variety of physical conditions, so we may have a variety of accretion regimes. Existing theories describe different regimes of black hole accretion flows, which can be realized under different physical conditions. Accreting black holes in nearby galactic nuclei and
low-state X-ray binaries are much dimmer than the standard Shakura-Sunyaev disc model would predict. A phenomenon of under-luminous accreting black holes in X-ray binaries and super-massive black holes in galactic nuclei has stimulated the recent investigations of radiation inefficient accretion flows (RIAFs)
(see Narayan, Mahadevan \& Quataert 1998, Kato, Mineshige \& Fukue 2008 for reviews). In such a flow, radiative losses are small because of low particle density
of accreting flow at low accretion rates. Contrary to standard Shakura-Sanyaev disc model, which successfully explains radiation soft and luminous X-ray sources, models of RIAFs are used to explain significant deficit of radiation observed in some X-ray sources. A particular example of such under-luminous sources is the Galactic center, Sagittarius $A^{*}$, with host a $2\times10^6 $ solar mass black hole. The Galactic center has a luminosity that is well bellow the estimated value based on Shakura-Sanyaev accretion disc Model (Melia \& Falcke 2001).

ADAFs have an opposite regime to that of the standard model. In the standard model, the flow is described
in such a way that the heat generated by the viscosity radiates out of the system immediately after its generation (Shakura \& Sunyaev
1973). These advection-dominated accretion flows occur in two regimes depending on their mass accretion and optical depth. Actually, the optical depths of accretion flows are highly dependent on their accretion rates. In a high mass-accretion rate, the optical depth becomes very high and the radiation generated by the accretion flow can be trapped within the disc.  In this case, the optical depth is very large, and photons, which carry the most of internal energy, are trapped inside the inflowing matter and can not be radiated away. This type of model is called `slim disc', or optically thick ADAF. Although the radiative efficiency of optically thick ADAF is also low, we usually only call the optically thin ADAF as RIAF. In the limit of low mass-accretion rate, the disc becomes optically thin. In this case, the cooling time of accretion flows is longer than the accreting time-scale. The energy
generated by accretion flows therefore remains mostly in the discs, and the discs cannot radiate their energy efficiently. This type of accretion
flow is named a radiation-inefficient accretion flow (RIAF). This type of accretion flow has been investigated by many authors (Narayan \& Yi 1994; Abramowicz et al. 1995; Chen 1995). At the same time as the ADAF model was proposed, it was realized the ADAFs are likely to be convectively unstable in the radial direction because of the inward increase of the entropy of accreting gas (Igumenshchev \& Abramowicz (1999), Stone, Pringle \& Begelman (1999), Igumenshchev, Narayan \& Abramowicz  (2003)). Most recent work focusing on the convective instability is by Yuan \& Bu (2010). Two and three dimensional simulations of low-viscosity RIAF have confirmed the convective instability in these flow (Igumenshchev, Chen \& Abramowicz 1996; Igumenshchev \& Abramowicz 2000, McKinney \& Gammie 2000). Narayan, Igumenshechev \& Abramowicz (2000) and Quataert \& Gruzinov (2000) construct another analytical model of RIAFs which was based on a self-similar solution which was called the convection dominated accretion flow (CDAF). CDAF consist of a hot plasma at about virial temperature and have a flattened time-averaged radial density profile, $\rho\propto r^{-1/2}$, where $\rho$ is the density and R is the radius. In CDAFs the most of energy which released in the innermost region of accretion flow is transport outward by convection motion.

The self-similar CDAF model as the same as other self-similar models is very clear and instructive, but it has some limitations. It is only a local, not a global solution for RIAF, it the sense that it can only be valid for a region far from boundaries. So it can not reproduce physical behavior of accretion flow it it's transonic radial motion- the most fundamental feature of black hole accretion flow. Abramowicz et al (2002) did suggest two-zone structure for RIAF: an outer convection dominated zone and an inner advection dominated zone separate at a transition radius $\sim 50 r_g$.

There is some observational evidence that accretion process is often and perhaps always associated with wind. Mass loss or wind appears
to be a common phenomenon in many astrophysical systems. These wind and outflow mechanisms are observed in micro-quasars,
YSOs ( Ferrari 1998, Bally, Reipurth \& Davis 2007; Whelan et al 2005). It is belief that disc wind/outflow contributes to loss of mass, angular momentum,
and thermal energy from accretion discs (e.g. Piran 1977; Blandford \& Payne 1982; Foschini 2011, Knigge 1999).
Various driving sources are proposed, such as thermal, radiative and magnetic
mechanism. The name of the wind depends on its driving
mechanism. In this paper we will follow the hydrodynamical (thermal)
wind which has been discussed by many authors (e.g. Meier 1979,
1982; Fukue 1989; Takahara, Rosner \& Kusnose 1989).

In this paper we will discuss the properties of CDAFs in a general large-scale magnetic field with hydrodynamical wind. We will
concentrate on the self-similar solution. This study is motivated by recent works of Zhang \& Dai 2008 who showed that the effect of large scale
magnetic field on the CDAFs with a constructive self-similar solution. Because of CDAFs usually modeled for outer regions of RIAFs where the
hydrodynamical outflow becomes important, we investigate the role of outflow in the dynamical structure of magnetized CDAFs.
We will show the basic equations and self-similar solution in next sections.

\section{Basic Equations}
\label{sect:Obs}

We are interested in analyzing the structure of a magnetized hot accretion flows
bathed in a global magnetic field where convection and wind play
an important role in energy and angular momentum transportation.
So we suppose a rotating and accreting disc around a compact
Schwarzchild black hole of mass $ M_* $. Thus, for a steady axi-symmetric
accretion flow, i.e., $ \partial/\partial t = \partial/\partial \varphi=0  $,
we can write the standard equations in the cylindrical
coordinates $ (r, \varphi, z) $ centered on the  accreting object. We vertically integrated
the flow equations and, all the physical variables become
only function of the radial distance $ r $. Moreover, we consider a magnetic
field in the disc with three components, $ B_r $, $ B_{\varphi} $ and $ B_z $.
We also neglect the relativistic effects and Newtonian gravity in radial
direction is considered.

The equation of continuity will be
\begin{equation}\label{continuity1}
    \frac{\partial}{\partial r}(r \Sigma v_r)+ \frac{1}{2 \pi}\frac{\partial \dot{M}_w}{\partial r} = 0
\end{equation}
where $ \Sigma $ is the surface density at the cylindrical radius $ r $,
which is define as $ \Sigma = 2 \rho H $, $ \rho $ being
midplane density and $ H $ the disc half-thickness and  $ v_r $
is the radial infall velocity. Also the mass loss rate by outflow/wind
is represented by $ \dot{M}_w $, so
\begin{equation}\label{Mdot1}
    \dot{M}_w(r) = \int 4 \pi r^{\prime}\dot{m}_w(r^{\prime})d r^{\prime}
\end{equation}
where $ \dot{m}_w(r) $ is the mass loss rate per unit area from each
disc face. We can write the dependence of accretion rate as follows
(Blandford \& Begelman 1999; Shadmehri 2008)

\begin{equation}\label{Mdot2}
    \dot{M} = -2 \pi r \Sigma v_{r} = \dot{M}_{0}(\frac{r}{r_{0}})^{s}
\end{equation}
where $ \dot{M}_{0} $ is the mass accretion rate at the outer radius of the disc $ r_{0} $ and $ s $ with s being constant of order unity (Blandford \& Begelman 1999).

Considering equations (\ref{continuity1})-(\ref{Mdot2}), we
can write
\begin{equation}\label{mdot_wind}
    \dot{m}_{w} = s \frac{\dot{M}_{0}}{4 \pi r^{2}_{0}} (\frac{r}{r_{0}})^{s - 2}
\end{equation}

The equation of motion in the radial direction is
\begin{equation}\label{motion_r}
	 v_r \frac{dv_r}{dr} = \frac{v^2_{\varphi}}{r} - \frac{GM_*}{r^2} - \frac{1}{\Sigma}\frac{d}{dr}(\Sigma c^2_{s}) - \frac{1}{2\Sigma}\frac{d}{dr}(\Sigma c^2_{\varphi} + \Sigma c^2_{z})  - \frac{c^2_{\varphi}}{r}
\end{equation}
where $ v_{\varphi} $ is the rotational velocity, $ c_s $ is the
sound speed, which is define as $ c^2_{s} \equiv p_{gas}/\rho $, $ p_{gas} $
being the gas pressure. Here, $ c_r $, $ c_{\varphi} $ and
$ c_z $ are Alfv$\acute{e}$n sound speeds in three direction of cylindrical
coordinate and define as
\begin{equation}
 c^2_{r, \varphi, z} =\frac{B^2_{r, \varphi, z}}{4 \pi \rho}= \frac{2 p_{{mag}_{ r, \varphi, z}}}{\rho}
\end{equation}

where $ p_{{mag}_{ r, \varphi, z}}  $ are the magnetic pressure in three directions.
The angular transfer equation with considering outflow/wind  and convection can be written as
\begin{equation}\label{motion_phi}
   \Sigma v_r \frac{d}{dr}(r v_{\varphi}) = - \frac{1}{r}\frac{d}{dr}(J_{vis}) - \frac{1}{r}\frac{d}{dr}(J_{con})+ r\sqrt{\Sigma} c_r \frac{d}{dr}(\sqrt{\Sigma} c_{\varphi}) +\Sigma c_r c_{\varphi}\\
    - \frac{l^2 (r\Omega)}{2 \pi}\frac{d \dot{M}_w}{dr}
\end{equation}

where $ J_{vis} $ and $ J_{con} $ are viscous and
convective angular momentum fluxes respectively which are defined as
\begin{equation}\label{j_vis}
J_{vis} = - \nu \Sigma r^3 \frac{d\Omega}{dr}
\end{equation}
and
\begin{equation}\label{j_con}
	J_{con} = - \nu_{con}\Sigma r^{3(1 + g)/2}\frac{d}{dr}(\Omega r^{3(1 - g)/2})
\end{equation}
Here, $ \nu $ is the kinematic viscosity coefficient, $ \nu = \alpha c_{s} H $, with $ \alpha $
being the constant Shakura \& Sunyaev parameter, $ \nu_{con} $ is
the convective diffusion coefficient, and $ g $ is the index to determine the condition of
convective angular momentum transport. When $ g = 1 $, the flux of angular momentum
due to convection is
\begin{equation}
	J_{con} = - \nu_{con} \Sigma r^3 \frac{d\Omega}{dr}
\end{equation}
and when $ g = -1/3 $, the convective angular momentum
flux will be (Narayan et al. 2000)
\begin{equation}
J_{con} = - \nu_{con} \Sigma r \frac{d( r^{2}\Omega)}{dr}
\end{equation}

The last term on the right side of angular transfer equation, eq (\ref{motion_phi}),
represents angular momentum
carried by the outflowing material(see e.g., Knigge 1999).

By integrating over $ z $ of the hydrostatic balance, we have
\begin{equation}\label{motion_z}
    \Omega^2_{K} H^2 - \frac{1}{\sqrt{\Sigma}}c_{r} \frac{d}{dr}(\sqrt{\Sigma} c_{z})H -\big[ c^2_{s} + \frac{1}{2}(c^2_{r} + c^2_{\varphi})\big]=0
\end{equation}

By considering outward energy due to convection and energy loss by outflows, the energy equation will be
\begin{equation}\label{energy1}
	\Sigma v_{r} T \frac{dS}{dr} + \frac{1}{r}\frac{d}{dr}(r F_{con}) = f (\nu +g \nu_{con})\Sigma r^{2}(\frac{d\Omega}{dr})^{2} - \frac{1}{2}\eta \dot{m}_{w}(r)v^{2}_{K}(r)
\end{equation}
in above equation $ T $ is temperature, $ S $ is the specific entropy and  $ F_{con} $ is the convective energy flux
which is defined as
\begin{equation}\label{convective energy}
	F_{con} = - \nu_{con} \Sigma T \frac{dS}{dr}
\end{equation}
where
\begin{equation}
T \frac{dS}{dr} = \frac{1}{\gamma -1 }\frac{dc^{2}_{s}}{dr} - \frac{c^{2}_{s}}{\rho}\frac{d\rho}{dr}
\end{equation}
here $ \gamma $ is the radio of specific heats.  As it mentioned, the last term on
right hand side of energy equation is the energy loss due to wind or
outflow (Knigge 1999). Depending on the energy loss mechanism,
dimensionless parameter $ \eta $ may change. In our case, we consider
it as a free parameter (Knigge 1999). Also in energy equation we still
neglect the Joule heating rate.

We adopt the assumptions of Narayan et al. (2000) and Lu et al. (2004) for the convective diffusion coefficient, $ \nu_{con} $, which is defined as

\begin{equation} \label{nu1}
\nu_{con} = \frac{L^{2}_{M}}{4}(- N^{2}_{eff})^{1/2}
\end{equation}
Here  $ N_{eff} $ is the effective frequency of convective blobs and  $ L_{M} $ is the characteristic mixing length.
The effective frequency of convective blobs will be
\begin{equation}
N^{2}_{eff} = N^{2} + \kappa^{2}
\end{equation}
with $ N $  and  $ \kappa $ being the Brunt-V\"{a}is\"{a}l\"{a} frequency and epicyclic frequency respectively, which are defined as
\begin{equation}
	N^{2} = -\frac{1}{\rho}\frac{dp_{g}}{dr}\frac{d}{dr}\ln\big(\frac{p^{1/\gamma}_{g}}{\rho}\big)
\end{equation}
\begin{equation}
	\kappa^{2} = 2 \Omega^{2} \frac{d \ln (r^{2} \Omega)}{d \ln r}
\end{equation}

Also, the characteristic mixing length $ L_{M} $ in terms of the pressure scale height, $ H_{p} $, can be written as
\begin{equation}
	L_{M} = 2^{-1/4}l_{M}H_{p}
\end{equation}
and
\begin{equation}
	H_{p} = - \frac{dr}{d\ln p_{g}}
\end{equation}
where $ l_{M} $ is the dimensionless mixing length parameter
and its amount is estimated to be equal to $ \sqrt{2} $ in ADAFs
(Narayan et al. 2000; Lu et al. 2004). The convective diffusion coefficient can also
write in the form similar to usual viscosity of Shakura \& Sunyaev (1973),
\begin{equation} \label{nu2}
	\nu_{con} = \alpha_{con} c_{s} H
\end{equation}
where $ \alpha_{con} $ is a dimensionless coefficient that describes
the strength of convective diffusion. The $ \alpha_{con} $ coefficient
can be obtained by equations (\ref{nu1}) and (\ref{nu2})
\begin{equation}
	\alpha_{con} = \frac{L^{2}_{M}}{4 c_{s} H} (-N^{2}_{eff})^{1/2}
\end{equation}
Finally we can write the three components of induction equation,
$ (\dot{B}_{r}, \dot{B}_{\varphi}, \dot{B}_{z} ) $, to measure
the magnetic field escaping rate,
\begin{equation}\label{induction1}
    \dot{B}_{r} = 0,
\end{equation}
\begin{equation}\label{induction2}
    \dot{B}_{\varphi} = \frac{d}{dr}(v_{\varphi} B_{r} - v_{r} B_{\varphi}),
\end{equation}
\begin{equation}\label{induction3}
    \dot{B}_{z} = -\frac{d}{dr}(v_{r} B_{z}) - \frac{v_{r} B_{z}}{r}.
\end{equation}
where $\dot{B}_{r,\varphi, z}$ is the field scaping/creating rate due to magnetic instability or dynamo effect.
Now we have a set of MHD equations that describe the structure of magnetized CDAFs. The solutions to these equations are strongly correlated to viscosity, convection, magnetic field strength $\beta_{r,\varphi,z}$ and the degree of advection $ f $. We seek a self-similar solution for the above equations. In the next section we will present self-similar solutions to these equations.

\section{Self-Similar Solutions}
\label{sect:Self-Similar}
\subsection{Analysis}
We seek self-similar solutions of the above equations. Therefore, according to Narayan and Yi 1994, we can write
similarity solutions as
\begin{equation}\label{self_sigma}
    \Sigma(r) = c_{0} \Sigma_{0} (\frac{r}{r_{0}})^{s-\frac{1}{2}}
\end{equation}
\begin{equation}\label{self_vr}
    v_{r}(r) = - c_{1} \sqrt{\frac{GM_*}{r_{0}}}(\frac{r}{r_{0}})^{-\frac{1}{2}}
\end{equation}
\begin{equation}\label{self_phi}
    v_{\varphi}(r) =  c_{2} \sqrt{\frac{GM_*}{r_{0}}}(\frac{r}{r_{0}})^{-\frac{1}{2}}
\end{equation}
\begin{equation}\label{self_cs}
    c^{2}_{s}(r) =  c^{2}_{3} (\frac{GM_*}{r_{0}})(\frac{r}{r_{0}})^{-1}
\end{equation}
\begin{equation}\label{self_cmag}
    c^{2}_{r,\varphi,z}(r) =  \frac{B^{2}_{r,\varphi,z}}{4 \pi \rho} = 2 \beta_{r,\varphi,z} c^{2}_{3} (\frac{GM_*}{r_{0}})(\frac{r}{r_{0}})^{-1}
\end{equation}
\begin{equation}\label{self_H}
    H(r) = c_{4} r_{0}(\frac{r}{r_{0}})
\end{equation}
where constants $ c_{0} $, $ c_{1} $, $ c_{2} $, $ c_{3} $ and $ c_{4} $ will be determined
later from the main MHD equation. $ \Sigma_{0} $ and $ r_{0} $ are exploited to
write the equations in non-dimensional forms and the constants
$ \beta_{r,\varphi,z} $ measure the radio of
the magnetic pressure in three direction to the gas pressure, i.e.,
$ \beta_{r,\varphi,z} = p_{{mag}_{r,\varphi,z}/p_{gas}} $.

In addition, the field scaping/creating rate $\dot{B}_{r,\varphi, z}$ is assume to be a form of

\begin{equation}\label{self_Bdot}
    \dot{B}_{r,\varphi,z} = \dot{B}_{r0,\varphi0,z0} (\frac{r}{r_{0}})^{\frac{1}{2}(s - \frac{11}{2})}
\end{equation}
where $ \dot{B}_{r0,\varphi0,z0} $ are constant.

By substituting the above self-similar solutions in
the continuity, momentum, angular momentum, hydrostatic
balance and energy equation of the disc, we obtain the following
system of dimensionless equations to be solved for $ c_{0} $, $ c_{1} $,$ c_{2} $
$ c_{3} $ and $c_{4}$:
\begin{equation}\label{mdot}
    c_{0}c_{1} = \dot{m}
\end{equation}

\begin{equation}\label{self_motionr}
	-\frac{1}{2} c^2_{1} = c^2_{2}  - 1  - \big[(s - \frac{3}{2})+(s - \frac{3}{2})\beta_z + (s + \frac{1}{2})\beta_{\varphi} \big]c^{2}_3
\end{equation}

\begin{equation}\label{self_motionphi}
   (sl^{2} - \frac{1}{2}) \frac{\dot{m}}{c_{0}}c_{2} = -\frac{3}{2} (s + \frac{1}{2}) (\alpha + g \alpha_{con}) c_2 c_{3} c_{4} +(s +  \frac{1}{2}) c^{2}_3 (\beta_r \beta_{\varphi})^{1/2}
\end{equation}

\begin{equation}\label{self-c4}
    c_{4} = \frac{1}{2} c_{3}\bigg\{ \bigg[(s - \frac{3}{2})^2 \beta_r \beta_{z} c^2_3 + 4 (1 + \beta_r + \beta_{\varphi})\bigg]^{1/2} + c_{3}(s - \frac{3}{2})(\beta_r \beta_{z})^{1/2} \bigg\}
\end{equation}

\begin{equation}\label{self-energy}
\big( \frac{1}{\gamma - 1} +s - \frac{3}{2}\big) \bigg\{ (s - 1) \alpha_{con} c^{3}_{3} c_{4} + c_1 c^{2}_3 \bigg\} = \frac{9}{4} f c^2_{2} c_{3} c_{4} (\alpha + g \alpha_{con}) - \frac{1}{4} s \eta \frac{\dot{m}}{c_{0}}
\end{equation}

where $ \dot{m} $ is the non-dimensional mass accretion rate which is defined as

\begin{equation}
	\dot{m} = \frac{\dot{M}_{0}}{2 \pi r_{0} \Sigma_{0} \sqrt{GM_*/r_{0}}}
\end{equation}

 If we solve the self-similar structure of the magnetic field escaping rate , we will have:
\begin{equation}\label{inducton11}
    \dot{B}_{0r} = 0
\end{equation}

\begin{equation}
\dot{B}_{0\varphi} = \frac{1}{2}(s -\frac{7}{2}) \frac{GM_*}{r^{5/2}_{0}} c_{3}\sqrt{\frac{4 \pi c_{0}  \Sigma_0}{ c_{4} }} \big(c_2 \sqrt{\beta_r}+ c_1 \sqrt{\beta_{\varphi}} \big)
\end{equation}

\begin{equation}\label{inducton33}
     \dot{B}_{0z}  = \frac{1}{2}(s - \frac{3}{2})  c_1 c_{3}\frac{GM_*}{r^{5/2}_{0}}   \sqrt{\frac{4 \pi  \beta_{z} c_{0}  \Sigma_0}{ c_{4} }}
\end{equation}

We can solve these simple equations numerically and clearly just physical
solution can be interpreted. Without mass outflow and magnetic
field, i.e. $ s =  l = \eta = \beta_{r} = \beta_{\varphi} = \beta_{z} = 0 $,
the equations and their similarity solutions are reduced to the
 Narayan et. al solution (Narayan \& Igumenshchiv \& Abramowicz 2000). Also in the absence of outflow
they are reduced to Zhang \& Dai 2008.

Now we can analysis behavior of the solutions in the presence of wind, convection
and global magnetic field. The parameters of our
model are the standard viscose parameter $ \alpha $, the advection
parameter $ f $, the radio of the specific heats $ \gamma $, the mass-loss
parameter $ s $, the degree of magnetic pressure to the gas pressure in
three dimensions of cylindrical coordinate, $ \beta_{r} $,
$ \beta_{\varphi} $ and $ \beta_{z} $ and $ l, \eta $ parameters corresponding to the rotation and
non-rotation wind and energy loss by wind.

\subsection{Numerical Results}

\begin{figure}[h!!!]
   \centering
   \includegraphics[width=15.0cm, angle=0]{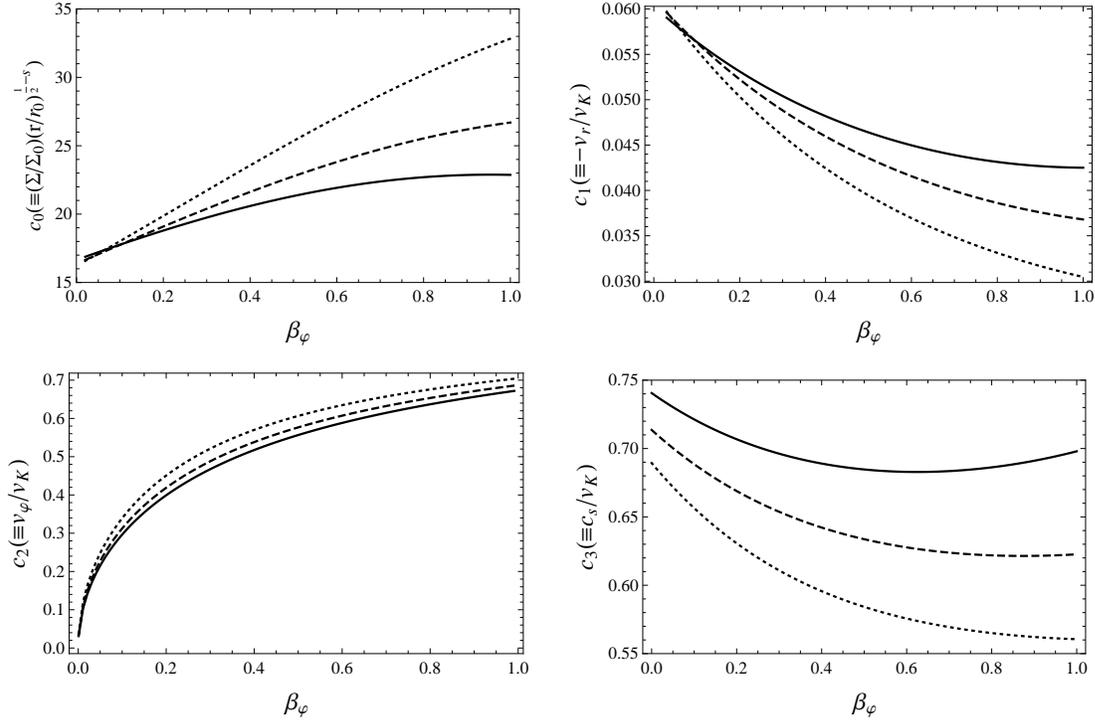}
   \caption{ Numerical coefficient $ c_{i} $s as a function of magnetic parameter
$ \beta_{\varphi} $ for several values of $ s $. The dotted, dashed and solid
lines correspond to $ s = 0.0, 0.1 $ and $ 0.2$ respectively. Parameters
are set as $ \alpha = 0.5 $, $ \gamma = 1 $, $ g = -1/3 $, $ \beta_{r} = \beta_{z} = 0.4 $, $ \eta =l = 1 $ and $ f = 1 $. }
   \label{beta_phi}
   \end{figure}

In Figure \ref{beta_phi}, the surface density $ (c_{0}) $, the radial
infall velocity $ (c_{1}) $, the rotational velocity $ (c_{2}) $ and the sound speed $ (c_{3}) $
are shown as a function of toroidal magnetic field parameter $ \beta_{\varphi} $ for several
values of wind parameter $ s $, i.e., $ s = 0 $ (dotted line, no wind), $ s = 0.1 $ (dashed line) and
$ s = 0.2 $ (solid line). Four panels of figure \ref{beta_phi} are set as $ \alpha = 0.5 $, $ \gamma = 1 $,
, $ \beta_{r} = \beta_{z} = 0.4 $, $ \eta = l = 1 $ and $ f = 1 $. by adding the toroidal magnetic field parameter $ \beta_{\varphi} $, we see the surface density and rotational velocity of the disc increase,
although the radial and sound speed both decrease.

On the other hand the radial flow decreases when the toroidal magnetic
field becomes large.
We can see that by adding the magnetic field influences (adding $ \beta_{\varphi} $), the rotation velocity will increase. This is because the disc should rotate faster than the case without the magnetic field which results in the magnetic tension.

 In figure \ref{beta_phi} we also studied the effect of
parameter $ s = 0 $ on physical coefficients. As it mentioned, the value of $ s $ measures the strength of
wind/outflow and a longer $ s $ denotes a stronger wind. We can see that for stronger outflow $ (s = 0) $,
the reduction of surface density is more evidence. We see that the convective model of accretion flows
with the presence of wind rotate more slowly than those without winds and wind leads to enhance accretion
velocity. Also the sound speed of the disc increase for stronger outflows.

\begin{figure}[h!!!]
   \centering
   \includegraphics[width=15.0cm, angle=0]{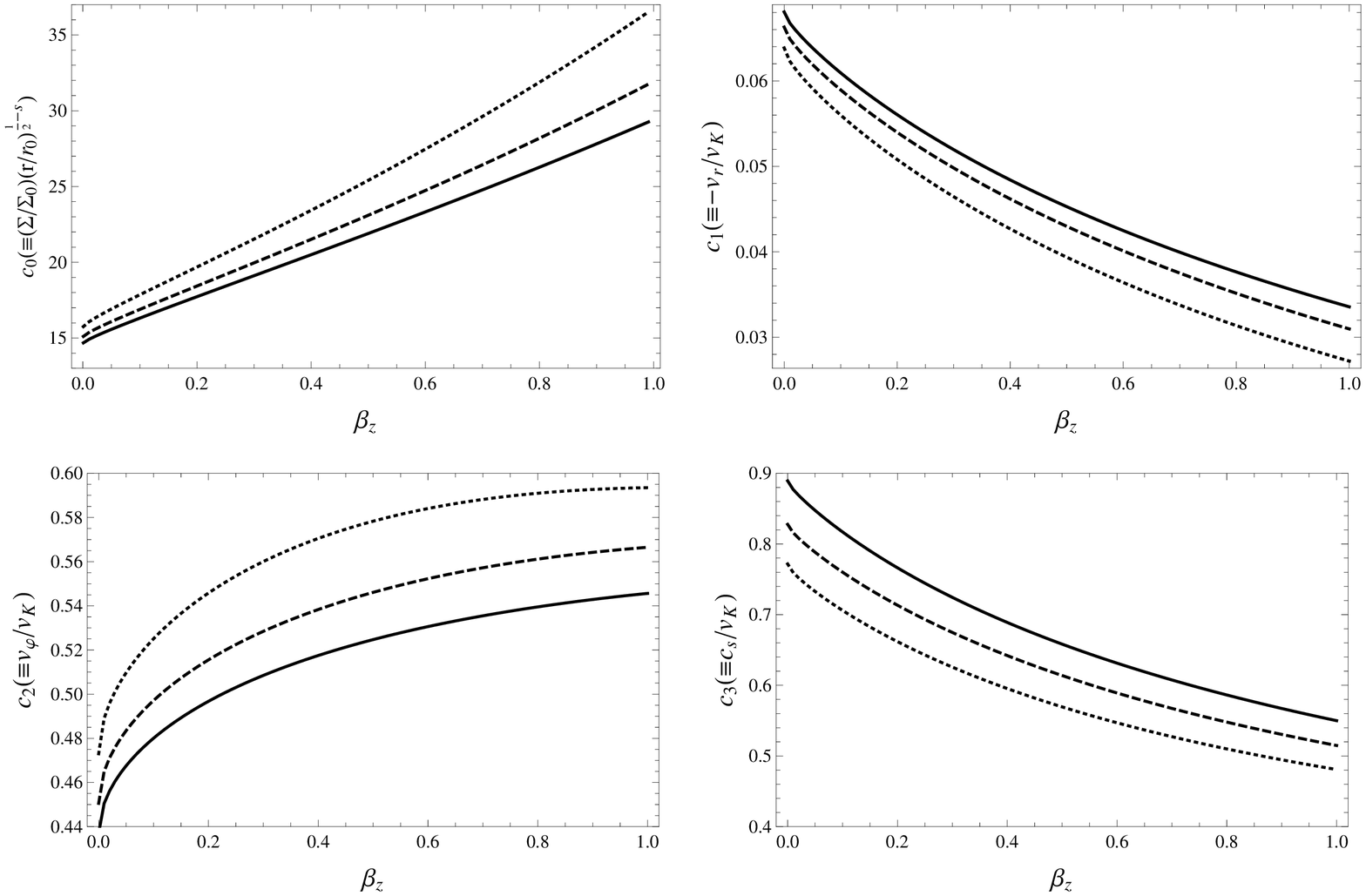}
   \caption{ Numerical coefficient $ c_{i} $s as a function of magnetic parameter
$ \beta_{z} $ for several values of $ s $. The dotted, dashed and solid
lines correspond to $ s = 0.0, 0.1 $ and $ 0.2$ respectively. Parameters
are set as $ \alpha = 0.5 $, $ \gamma = 1 $, $ g = -1/3 $, $ \beta_{r} = \beta_{\varphi} = 0.4 $, $ \eta =l = 1 $ and $ f = 1 $.  }
   \label{beta_z}
   \end{figure}

Our variables as a function of magnetic field parameter $ \beta_{z} $ and several values of mass loses are shown in Figure \ref{beta_z}. As it is seen, a strong $ z $-component of magnetic field leads to increase
both surface density and rotational velocity, while the radial infall velocity of materials and sound speed of the disc decrease.

\begin{figure}[h!!!]
   \centering
   \includegraphics[width=15.0cm, angle=0]{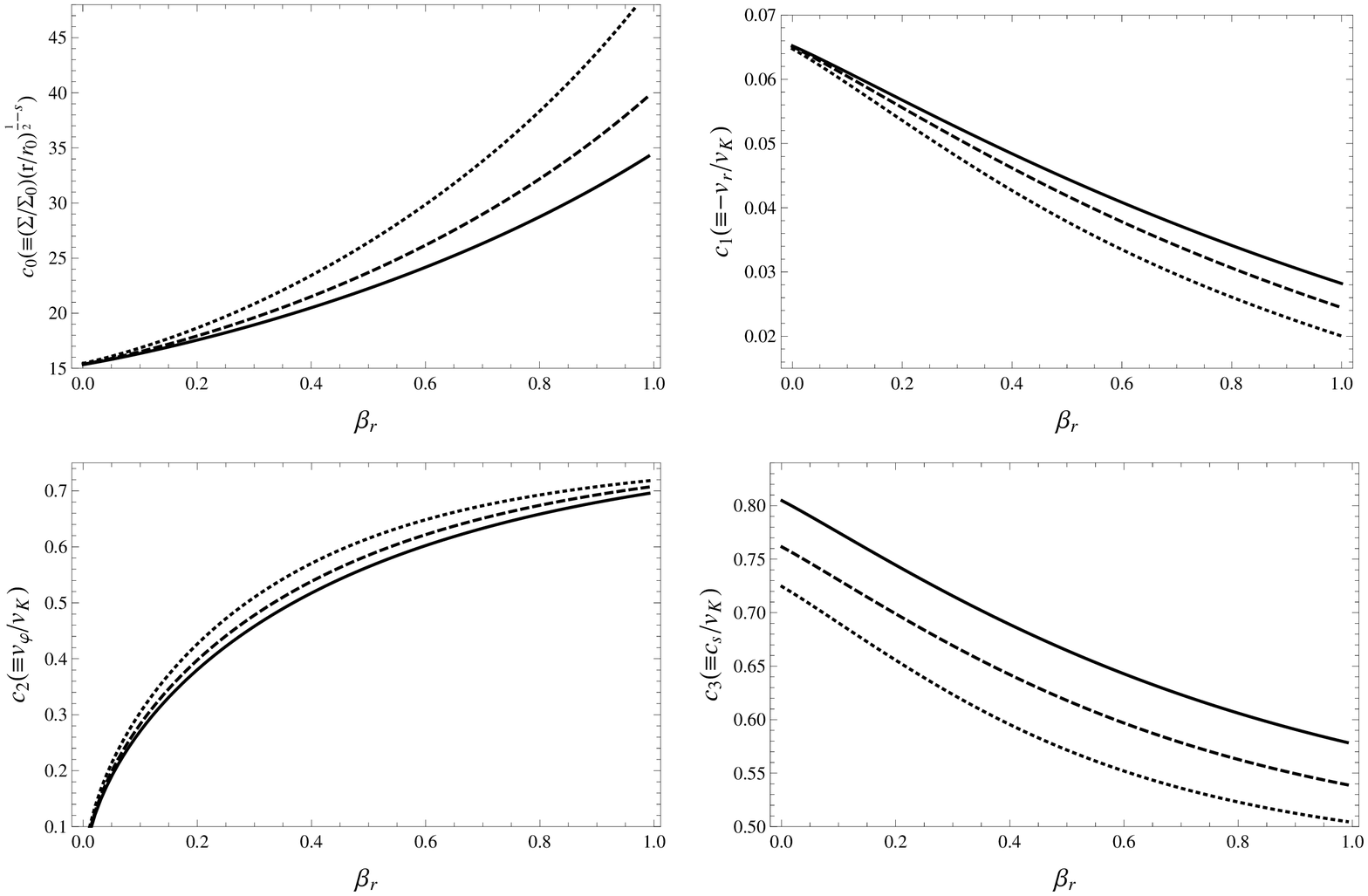}
   \caption{ Numerical coefficient $ c_{i} $s as a function of magnetic parameter
$ \beta_{r} $ for several values of $ s $. The dotted, dashed and solid
lines correspond to $ s = 0.0, 0.1 $ and $ 0.2$ respectively. Parameters
are set as $ \alpha = 0.5 $, $ \gamma = 1 $, $ g = -1/3 $, $ \beta_{\varphi} = \beta_{z} = 0.4 $, $ \eta =l = 1 $ and $ f = 1 $. }
   \label{beta_r}
   \end{figure}

Figure \ref{beta_r} shows how the coefficients $ c_{i} $s depend on the magnetic field parameter in radial
direction $ \beta_{r} $ for several values of outflow parameter $ s $. we see that the surface density and
rotational velocity raise up when the magnetic field parameter $ \beta_{r} $ increase. While the sound speed
and accretion velocity will decrease. According to figures \ref{beta_phi}, \ref{beta_z} and \ref{beta_r},
by adding all components of magnetic field, the surface density and rotational velocity increase although the
sound speed and radial infall velocity of the disc will decrease.

\begin{figure}[h!!!]
   \centering
   \includegraphics[width=16.0cm, height=45mm,angle=0]{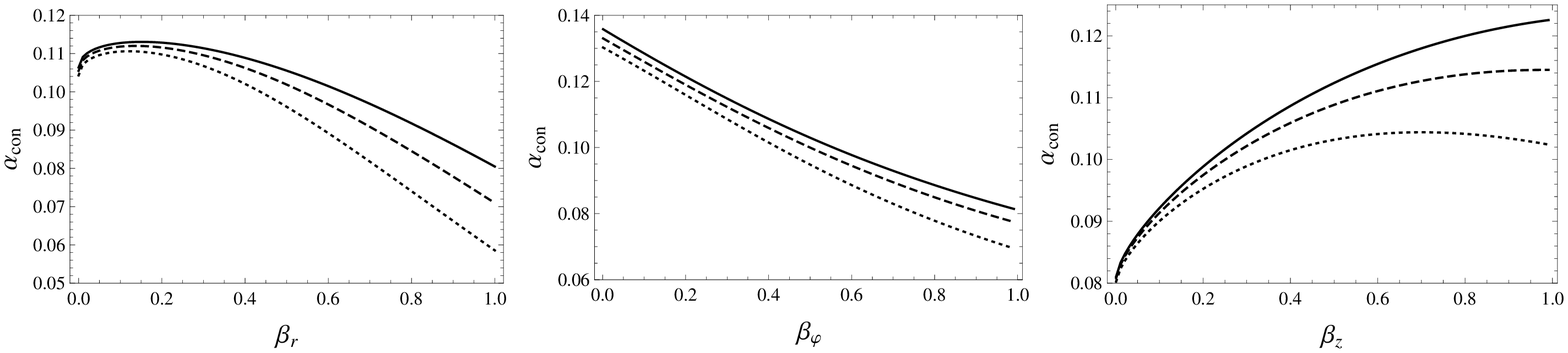}
   \caption{ The convective coefficient $ \alpha_{con} $ as a function of magnetic parameters $ \beta_{r} $, $ \beta_{\varphi} $ and $ \beta_{z} $ for several values of wind parameter $ s $. The dotted, dashed and solid
lines correspond to $ s = 0.0, 0.1 $ and $ 0.2$ respectively. Parameters
are set as $ \alpha = 0.5 $, $ \gamma = 1 $, $ g = -1/3 $, $ \beta_{r}=  \beta_{\varphi} = \beta_{z} = 0.4 $, $ \eta =l = 1 $ and $ f = 1 $.  }
   \label{alphacon}
   \end{figure}

In figure \ref{alphacon} we have plotted the convective parameter $ \alpha_{con} $ versus magnetic
field parameters $ \beta_{r} $, $ \beta_{\varphi} $ and $ \beta_{z} $ for several values of $ s $.
As it is seen, when the magnetic field parameter $ \beta_{r} $ become stronger, the convective parameter
decrease (left panel). Also the toroidal component of magnetic field $ \beta_{\varphi} $ decrease the $ \alpha_{con} $ too (middle panel), although  by adding $ z $-component of magnetic field the convective parameter $ \alpha_{con} $ increase (right panel).

\begin{figure}[h!!!]
   \centering
   \includegraphics[width=16.0cm, height=45mm,angle=0]{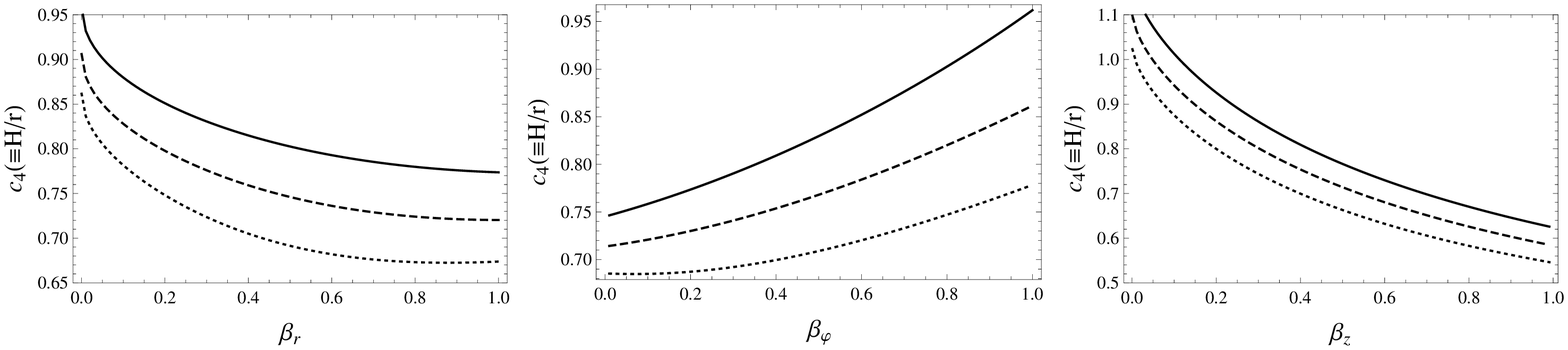}
   \caption{ The bihavior of $ H/r $ as a function of magnetic parameters $ \beta_{r} $, $ \beta_{\varphi} $ and $ \beta_{z} $ for several values of wind parameter $ s $. The dotted, dashed and solid
lines correspond to $ s = 0.0, 0.1 $ and $ 0.2$ respectively. Parameters
are set as $ \alpha = 0.5 $, $ \gamma = 1 $, $ g = -1/3 $, $ \beta_{r}=  \beta_{\varphi} = \beta_{z} = 0.4 $, $ \eta =l = 1 $ and $ f = 1 $.  }
   \label{thickness}
   \end{figure}

Figure \ref{thickness} represent that the disc thickness enhance with toroidal magnetic field parameter
$ \beta_{\varphi} $ and wind parameter $ s $ (middle panel). While the radial and vertical components of magnetic field decrease the vertical thickness of the accretion disc (left and right panels).

\section{Conclusions}
\label{sect:conclusion}
The CDAF model consistently represent radiatively inefficient accretion flows into black holes in
the framework of viscose flow. We have considered the radial structure of RIAFs based on a self-similar analysis (CDAF).
Some approximations were made in order to simplify the main equations.
We assumed an axially symmetric, static disc with the $\alpha$-prescription of viscosity, $ \nu = \alpha c_{s} H $.
A set of similarity solution was presented for such a configuration.
We have extended Akizuki \& Fukue 2006; Zhang \& Dai 2008 and Faghei 2012 self
similar solutions to present dynamical structure of the convection dominated
accretion flows. We ignored the relativistic effects and the self-gravity of the discs.

We have found out that by increasing all components of magnetic field , the surface density and
rotational velocity increase although the sound speed and radial infall velocity of the disc decrease.
Also we have shown that the existence of wind will lead to reduction of surface density as well as rotational velocity. Moreover the radial velocity, sound speed, advection parameter and the vertical thickness of the disc will increase when outflow becomes important in the radiation inefficiently accretion flow.

In this manuscript we have follow the effect of Large-scale B-field on the structure convective this with wind and outflow. in the future it would be interesting to study how these effect would change the observational appearance of the flow.

Although we have made some simplifying assumptions in order to treat
the problem analytically, our self-similar solution shows
Large-scale magnetic fields can really change typical behavior of the
physical quantities of a hot accretion flow. Not only the
surface density of the disc changes, but also the rotational
and the radial velocities significantly change because of the magnetic fields.
It means any realistic model for hot disc should consider the possible effects of the
magnetic fields.

\label{lastpage}

\end{document}